\documentclass[aps,twocolumn,superscriptaddress,showpacs]{revtex4}
\usepackage{epsfig}
\usepackage{changebar}
\usepackage{graphicx}
\usepackage{epsfig}
\usepackage{subfigure}
\usepackage{amsmath}
\usepackage{rotating}
\usepackage{amsfonts}
\usepackage{amssymb} 
\newlength {\oldtextheight}
\newlength {\oldheadsep}
\psfigdriver{dvips}

\begin{document}
\title{Tempo and Mode of Evolution in the Tangled Nature Model}  
\author{Dominic Jones and Henrik Jeldtoft Jensen}
\email{dominic.jones@imperial.ac.uk}
\email{h.jensen@imperial.ac.uk}
\affiliation{Institute of Mathematical Sciences, 53 Princes' Gate, Imperial College London, London SW7 2PG \& Department of Mathematics, Imperial College London, South Kensington campus, London SW7 2AZ, UK}
\author{Paolo Sibani}
\affiliation{Institut for Fysik og Kemi, SDU, DK5230 Odense M, Denmark} 
\email{paolo.sibani@ifk.sdu.dk}
\date{15.3.2009} 
\begin{abstract}  
We study the Tangled Nature model of macro evolution and demonstrate that the co-evolutionary dynamics produces an increasingly correlated core of well occupied types. At the same time the entire configuration of types becomes increasing de-correlated. This finding is related to ecosystem evolution. The systems level dynamics of the model is subordinated to intermittent transitions between meta-stable states. We improve on previous studies of the statistics of the transition times and show that the fluctuations in the offspring probability decreases with number of transitions. The longtime adaptation, as seen by an increasing population size is demonstrated to be related to the convexity of the offspring probability. We explain how the models behaviour is a mathematical reflection  of Darwin's concept of {\em adaptation of profitable variations}.  
\end{abstract} 
\maketitle

\section{Motivation}  

In his 1944 book~\cite{Simpson1944},  Simpson's expressed aim was to reconcile the fruits of paleontological research 
with the Darwinian synthesis. The Darwinian synthesis, such as it was, consisted of Darwin's original ideas concerning 
competition for scarce resources and natural selection married to more recent discoveries revealing the first genetic,
 then fundamentally molecular, mechanisms by which Darwin's ideas were realised.  The question that Simpson wanted to 
 answer was: is it possible to infer anything about the \emph{mode} of evolution simply by observing its tempo.

The point, that has been made in different ways since Simpson's seminal work, is that, as Simpson has it, 
and as Gould has subsequently underlined \cite{Gould1995}, 
what holds for ten rats over a hundred years may not be true, or may not be simply 
extrapolated, to a million rats over a billion years. 
In this paper, we take this point further, by observing the tempo of evolution of a model of 
 interacting populations of different species. We find that there are two separate dynamical regimes on 
 which evolution takes place. In one  of those, the tempo  gradually decelerates  due to 
  the frustration in the interactions between different species. In the other,  observable quantities
  fluctuate around stationary values. 

A related issue is whether evolutionary dynamics on all spatial scales, ranging from  bacterial colonies confined
 in a flask (see eg.~\cite{Lenski2008})  to  biological species can be understood, at least in an approximate fashion,
  using  a small set of unifying theoretical principles. These principles would explain the emergent
   properties of evolutionary systems, for example the decrease of the rate of evolutionary events  
   on coarse-grained time scales~\cite{Raup1982}, in spite of  the microscopic rate of change remaining 
   constant. Clearly, evolution at the collective level is a non-stationary process, characterised by various trends.
    How are these trends related to the dynamics at the level of individuals? .

An approach  which has gained momentum due to the increasing speed and availability of computers,
is to search for unifying principles by simulating simple evolutionary models with  Monte Carlo 
techniques. These models blend a random element, the noise, with deterministic rules  for the average 
rates of reproduction, mutation and death in a group or groups of individuals. 
 Precursors to our approach include Kauffman's NK(C) model, which has been used primarily as a tool 
 to study the way fitness landscapes affect the course of evolution \cite{Kauffman1990}. 
 Our approach, while partially derived from the NKC model, is different in two ways: 
 we have frequency dependent interaction, and we do not explicitly build in a optimisation. 
 The fact that our model nevertheless appears to optimise certain macroscopic parameters is a matter of some interest, 
 and is discussed further below.

There are many elements of the puzzle that we will simply neglect. Indeed the model we shall 
study does not even have the rudimentary innovation of sexual reproduction. This approach, 
we should make clear, is limited and cannot provide a complete answer to the question of how the microscopic 
and macroscopic are related. However, we show that even such a simple model as the one discussed 
can give rise to macroscopic evolution of a very different tempo from the individual dynamics on which it is based, 
and we show how this emergent macroscopic behaviour comes about from the interaction of many individuals
 following Darwinian principles. It is the glassy nature of evolutionary dynamics - the existence of 
 quenched disorder giving rise to slow relaxation - 
 that in our model produces the macroscopic tempo of a different nature from that of the underlying dynamics.

\section{The Microscopic Level}
\label{sec:micro}
The Tangled Nature model of co-evolution \cite{Christ2002} has already been studied in 
several contexts. Its simplicity along with the rich complexity of its resulting 
behaviour makes it a paradigmatic model for testing co-evolutionary ideas. 
The model retains the binary string genotype geometry found in previous approaches, 
but replaces their `ad hoc'  static fitness  landscapes with a set of population  dependent 
interactions between extant species, similar to  the `tangled' interactions of an eco-system. 
This simple addition creates a fascinating complex behaviour:  the model retains the  intermittent 
dynamics found for adaptive walks in fitness landscapes\cite{Sibani99a}, where  periods of quasi-stasis 
which are long with respect to the timescale of the underlying, reproductive dynamics, 
are interspersed by  transition events, termed quakes, which  occur at a decreasing rate. 
The overall tempo of evolution is decelerating in the model,  indicating the possibility of an underlying   process optimising a hidden variable.
Importantly, the model  introduces new, and biologically relevant,  variables in the   dynamics at the macroscopic level. 
As the system evolves from a `random' initial state in the presence of scarce external resources,  
the network of extant and interacting population  changes,  slowly, but radically over time, enabling the system 
to support an ever growing number of individuals in the face of a constant supply of external resources.

Despite its simplicity, the model is able to reproduce the long time decrease reported in the overall macroscopic
 extinction rate, the observed intermittent nature of macro-evolution, denoted punctuated equilibrium 
 by Gould and Eldrdege~\cite{Gould1977}, the log-normal shape often observed for the species abundance 
 distributions~\cite{Anderson2004}, the power law relation often seen between area and the number
  of different species number~\cite{Jensen2006}, the frame work of the model is also able to reproduce 
  often reported exponential degree distributions of the network of species as well as the decreasing 
  connectance with increasing species diversity that has attracted much observational and theoretical interest~\cite{Laird2007}.      

Individuals are defined by a binary string of length $L$, and are thus points in an $L$ dimensional hypercube, whose elements, or nodes, will be denoted by Latin letters $i$ and $j$.
An \emph{a  priori} given fixed and  random   matrix $J_{ij}$, specifies  the strength and type of potential interaction between  populations at nodes $i$  and $j$. The evolutionary dynamics of the model, to be explained in detail below, leads to the occupancy of types evolving with time. Mutation allows new nodes  to become inhabited and existing occupied types to go extinct. The configuration of occupied types changes accordingly with time. Although the dynamics at the level of individuals is smooth and always happens at the same rate, at the systems level, i.e. the level of occupied configurations in type space, the dynamics is  highly intermittent, and, as mentioned, decelerating.   
The dynamics of the model is characterised by a succession of quasi stable states 
(or  QESS: quasi  evolutionary stable strategy). Each QESS is described by a 
network of sites which are populated and interact with each other. As the system undergoes the intermittent burst of changes to the occupancy one network of occupied sites is replaced by another.

The long time evolution of the occupancy in type space, and all the properties that can be derived form the composition and interactions 
of the extant population, 
are subordinated to the transitions, or the quakes, as they are customarily denoted in record dynamics. 
This is because no essential change occurs during the metastable state. The occupancy of individual sites may fluctuate but, 
since essentially the same set of types are present during the metastable periods,  no net drift can manifest itself. 
The most conspicuous overall effect of the succession of  QESS  is a logarithmically slow increase in the overall population size. 

The evolution of the configuration in type space is non-stationary nature and the population grows logarithmically with time as the configurations, 
through the adaptive search that takes place during the quakes events, manage to become collectively better  adapted. 
The adaptation consists in selecting networks of occupied sites in type space connected by increasingly mutualistic or beneficial 
interactions, which enables the temporal averaged net-reproduction rate to slightly overcome the killing rate.

A time-step in the model is carried out as follows: 
an individual is picked at random. Its offspring probability is calculated by first calculating the value of 

$$ H= \frac{C}{N}\sum_j J_{ij} n_j - \mu N $$

 where N is the total population, $n_j$ the site populations, $C$ a strength parameter and $\mu$ the carrying capacity 
 determining the resources available to the population. This value is then mapped on to the interval {0,1} using the function

$$P_{off}= \frac{1}{1+\exp(-H)},$$

which is the probability of reproduction in that time step. Reproduction is asexual: the parent is replaced by two copies of itself, 
with each offspring susceptible to mutations with probability $p_{mut}$ per genome site. In the next step a 
randomly chosen individual is killed with constant probability $p_{kill}$. Note that while a point mutation creates a new individual 
on a neighboring site of the hypercube on which populatons reside, the interaction matrix generally involves populations located
at arbitrary positons on  the hypercube.

Typically the   population is initially located on a randomly chosen, single site. 
To start with, the population decreases. Then, as reproduction becomes more likely, mutants begin to populate neighbouring sites. 
Soon a stable ``ecosystem'' is formed, a subset of genotypes whose mutual interactions form a (meta)stable dynamical system. 
This subsystem can be characterised by some key macroscopic properties - the average total number of individuals
in the system, the average number of sites occupied, and the properties of the realised interaction network.

\section{The Macroscopic Level}
\subsection{Centre of mass}
Since the system is made up of weighted nodes (each with a vector position in genotype space) interacting 
with each other, we can define a quantity analogous to the centre of mass in classical mechanics. 
We define this as the species occupancy weighted average position of the system in genotype space, $\bold{A}$, where
\begin{equation}
\bold{A}=\frac{1}{N}\sum_jn_j\bold{s}_j
\end{equation}
where $\bold{s}_j$ denotes the position of vertex number $j=1,...,2^L$ in the $L$ dimensional hypercube of genotypes, $n_j$ is the occupancy of species $j$ located at vertex $\bold{s}_j$
and $N$ is the total population: ($N=\sum_jn_j$).

During meta-stable periods this quantity remains practically constant, since it is only slightly modified by 
population fluctuations, which chiefly occur on the periphery of the system due to mutations. 
In contrast, when a quake occurs the centre of mass undergoes a major displacement. While it may be theoretically possible that a quake would occur which does not shift the centre of mass perceptibly, this is highly unlikely since any nearby stable configuration would undermine the stability of the current configuration.

\begin{figure}[htbp]
\centering
\includegraphics[width=10cm]{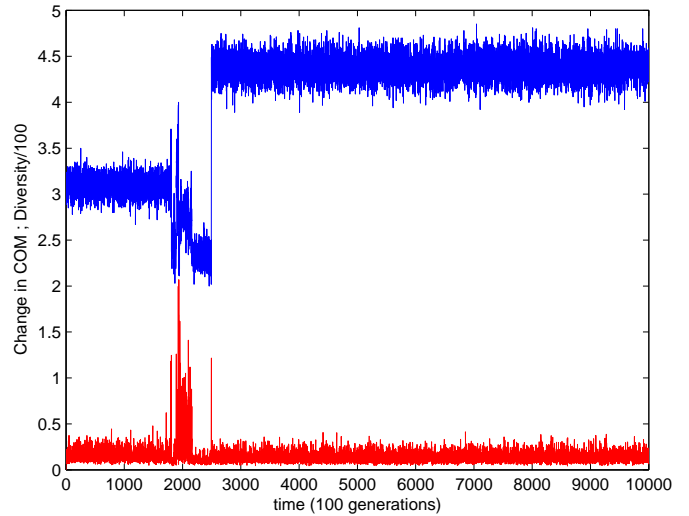}
\caption{The time evolution for a typical run of the Tangled Nature model. The diversity (blue) displays intermittent behaviour, exhibiting long plateaux interspersed with chaotic fluctuations (the diversity counts the number of species in the system, and behaves in much the same way as the population). The change in centre of the mass of the system (red) follows the same pattern - in general there is very little movement, with occasional big leaps in position. }
\label{fig:COM}
\end{figure}

The CM then   unambiguously identifies significant rearrangements in genotype space. 
In what follows, monitoring the CM position allows us to explore how the tempo of such rearrangements evolves
with the age of the system, counted from the beginning of the simulation..

Figure \ref{fig:COM} plots the changes in this quantity, with a plot of the total population for reference.
 Importantly we see that both variables exhibit the same behaviour, and it follows that one can be used as a good proxy for the other.
  In what follows we analyse population increments as a proxy for the underlying shift in configuration, since this is much more 
  computationally efficient.

\subsection{Aging}
There are two distinct tempos in the dynamics of the model, 
as has been suggested in previous papers. The nature of these two tempos is investigated here. 
Figure \ref{fig:COM} gives a qualitative picture of their origin - we see that the dynamics can be broadly divided 
into chaotic and stable phases, with the stable phases tending to be long-lived, as opposed to the chaotic phases which tend to be brief
 interludes. Within the stable phases there is still some dynamics due to the stochastic nature of the system: fluctuations
and sometimes natural cycles  around a fixed point are observed. Thus, there are two obvious timescales in the system - 
one governing the fluctuations within each stable phase, the other the chaotic transitions between stable phases.

To determine the tempo of these phases, or  more precisely to determine how the frequency of these two types of events scale
 with the age of the system we use a technique borrowed from analysis of spin glasses \cite{Sibani:2003p22,Sibani07}. 
 In spin glasses the thermally activated evolution can be analyzed by constructing the
 Probability Density Function (PDF) of the heat flow between the system and its thermal reservoir.
  Large, irreversible flows of heat out of the system
 only occur when the system undergoes a significant rearrangement, or quake. Consequently, observing how the PDF   evolves
  over time reveals  how the tempo of reversible and irreversible events evolve over time. 
  In our case, an exact  equivalent of the energy in a spin glass is not available.
Instead, we choose a characteristic scalar function of the dynamics, the population. 
In spin glasses the energy has a much more well defined role in the dynamics than does the population in the Tangled Nature model 
-possible changes in configuration are namely accepted or rejected based on energetic considerations, whereas the population plays no such 
discriminatory role in the present model. 

\begin{figure}[htbp]
\centering
\includegraphics[width=10cm]{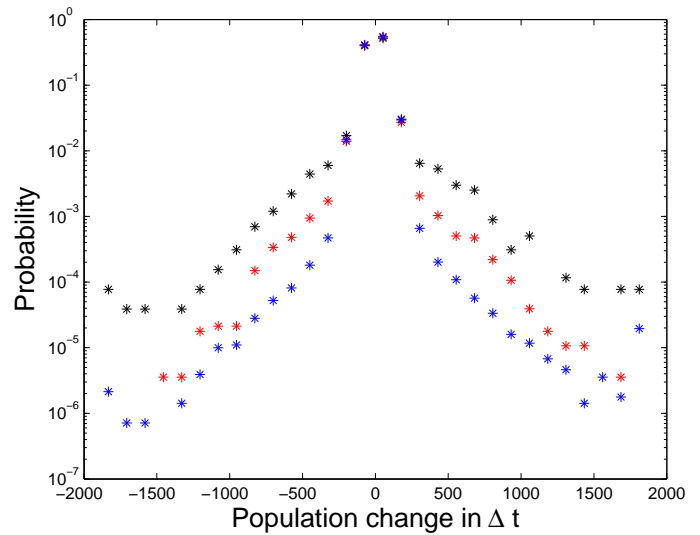}
\caption{The PDF of population increments for an ensemble of 1500 runs of $10^6$ generations. Black: $t_w = 1000$, red $t_w = 10000$, blue $t_w = 100000$. In this plot $\Delta t = 10$. The older the system, the less probable a large population fluctuation becomes.}
\label{fig:unsc}
\end{figure}

\begin{figure}[htbp]
\centering
\includegraphics[width=10cm]{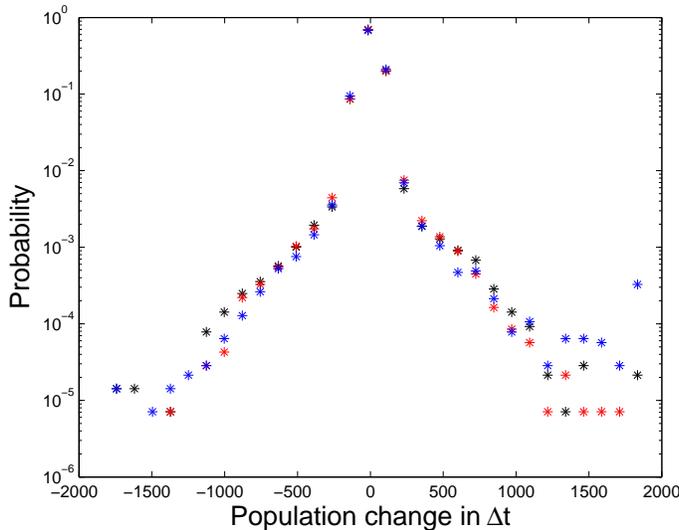}
\caption{The PDF of population increments for an ensemble of 1500 runs of $10^6$ generations. Black: $t_w = 1000$, red $t_w = 10000$, blue $t_w = 100000$. The total number of data points is equal for each age of the system; that is the sampling period $\Delta t = \frac{t_w}{100}$ scales with the age of the system.  The data collapses onto a single curve, showing that the probability of a large fluctuation scales like $\frac{1}{t_w}$}
\label{fig:sc}
\end{figure}

Letting  $t_w$ be the age of the system, we consider the distribution of population increments $$N(t_w +k\Delta t)-N(t_w + (k-1)\Delta t),$$  
where $k = 1,2,3,..., \frac{t_w}{\Delta t K}$, i.e.  at different epochs in the system's evolution over the observation
interval $[t_w, t_w + \frac{t_w}{K}]$, 
. The nature of the PDF will be determined by three parameters - $t_w$, $K$ and $\Delta t$.
The first quantity, $t_w$, measures  as mentioned the age of the system, while $K$ is inversely proportional to the size of the observation window. 
The last parameter is the number of Monte Carlo sweeps 
between consecutive measurements of the population size.
We need the observation window to be large enough to capture both the quasi-equilibrium and non-equilibrium dynamics, while not being 
so long that the system changes too significantly over the window. In practice this is a matter of trial and error. In figures \ref{fig:unsc} and \ref{fig:sc} the PDF of the population fluctuations is plotted for three different epochs, with $t_w = 1000, 10 000, 100 000$ and $K = 4$. The data comes from an ensemble of 1500 runs. 

Figure~\ref{fig:unsc} shows the data with $\Delta t =10$ for all epochs.The first thing to note is that the PDF has the characteristic 
shape common to similar spin glass studies, with one important difference. 
Since the data is plotted on linear-log axes, the parabolic shape in the centre represents a gaussian distribution, and the straight 
tails are exponential. The difference from the spin glass case is that here we have both positive and negative tails, 
whereas a in the heat flow PDF  for a spin glass the positive  tail is  undetectable, corresponding to the fact that   quakes
 reduce the energy of the system. The population, on the other hand, may go up or down in a quake, although, 
 as we shall see, there is still an overall tendency for the population to increase. 

In analogy with spin glasses, we identify the large, rare events in the tails with the irreversible quakes in the system.
 We see that as the system ages, these quakes become less and less frequent \emph{while the gaussian peak remains unchanged}. 
 We quantify this observation by rescaling the data - we anticipate a logarithmic decrease in the quake frequency, 
 and so we make $\Delta t$ proportional to the age of the system.  When we do this, the PDF from three different 
 ages collapses on to a single curve, so that the probability of large fluctuations (that is fluctuations appearing
  in the exponential tails) must scale as $t_w^{-1}$, since the probability of a large fluctuation in time $\Delta t$ 
  is to leading order $\Delta t P(t_w)$, where $P(t_w)$ is the probability of a large fluctuation as a function of $t_w$.


\begin{figure}[htbp]
\centering
\includegraphics[width=10cm]{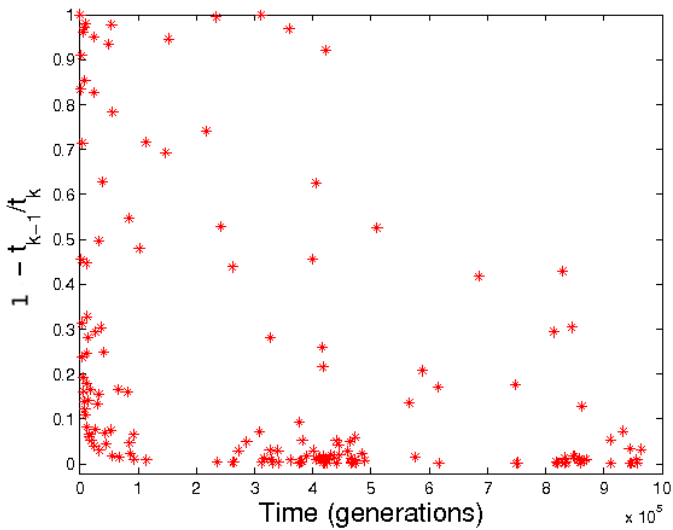}
\caption{ Plot of a sample of quake waiting time ratios, where $t_k$ is the time of the $k^{th}$ quake.  A Poisson distribution of quakes would give a more or less zero for long times, since the ratio would decay as $\frac{1}{\lambda t}$ for some average rate $\lambda$. Instead we see many non zero points for long times. The clustering of points along the $t$ axis is due to small 'secondary quakes' that are hard to distinguish from primary quakes. }
\label{fig:WT}
\end{figure}

The waiting time distribution for quakes (figure~\ref{fig:WT}) shows that the ratio of waiting time to age of the system remains non zero right throughout a million generation run, giving further proof that the quake probability decays with the age of the system.

\section{Discussion}

We have shown that by studying the tempo of the evolution in a schematic model we can infer a great deal about the mode - broadly speaking we find that we can split the dynamics into two bits, one quasi-equilibrium 
and the other non-equilibrium relaxation. Here we would like to suggest a mechanism for the observed slowing down of the dynamics 
at the macroscopic scale. First, let us recall the work done in spin glasses on similar aging phenomena. 
It has been found there that many microscopically different systems can be modelled by a unifying mesoscopic principle of barrier hopping. 
In this picture the system becomes increasing stable against quakes as each quake corresponds to the jumping of a barrier in the energy landscape
which is marginally larger than the barrier corresponding to the previous quake. 
Consequently, the evolution of the system is well described by record dynamics - each new maximum energy leads to a change in configuration 
and a new energy minimum. The energy landscape is such (for more detailed discussion see~\cite{Sibani:1998p8}) 
that each new well is deeper than the previous one, so the relaxation of the system slows as records become less and less frequent.

We propose that the similar  relaxation dynamics seen in the Tangled Nature model is due to a similar scenario, with some caveats.  
 One way of interpreting the spin glass dynamics is that the system falls into wells of increasing depth, from which it becomes 
 increasingly hard to escape since the magnitude of the fluctuations of the energy about each new fixed point remains constant. 
 In the tangled nature model, it is harder to justify an energy landscape picture, but we can measure the magnitude of fluctuations of the 
 birthrate, which is a measure of the systems ability to `break out' from each equilibrium state. To do this we have determined the standard deviation
  of the birth rate as a function of the age of the system, sampled during the different equilibrium stages
  of the dynamics. Indeed we find that there is a significant decrease in the fluctuation width over a simulation of $10^6$ generations.

To complete the circle, we also need to understand why a decrease in the birthrate probability fluctuations in turn leads to 
greater stability. A spike in the reproduction probability of the system has an effect beyond simple increase in numbers -
 it also increases the likelihood that new species in different parts of genotype space, generated by mutations, will interact 
 and thus destabilise the current configuration. It should be emphasised that the difference between the fluctuations 
 in the average probability and the fluctuations of the population amount to a difference in the \emph{weighting} of the core and the periphery
 of the system - that is between between vital, heavily populated genotypes and transient, sparsely populated ones. When considering the 
 birthrate, each site is weighted equally - so the fact that we see a decrease in these fluctuations while seeing a constant standard deviation 
 in the population is not contradictory -   it merely shows that the fluctuations are increasingly limited to sites, which are gradually becoming more similar in terms of the network of interactions $J_{ij}$ surrounding the sites carrying the larger proportion of the population.

\begin{figure}[htbp]
\centering
\includegraphics[width=10cm]{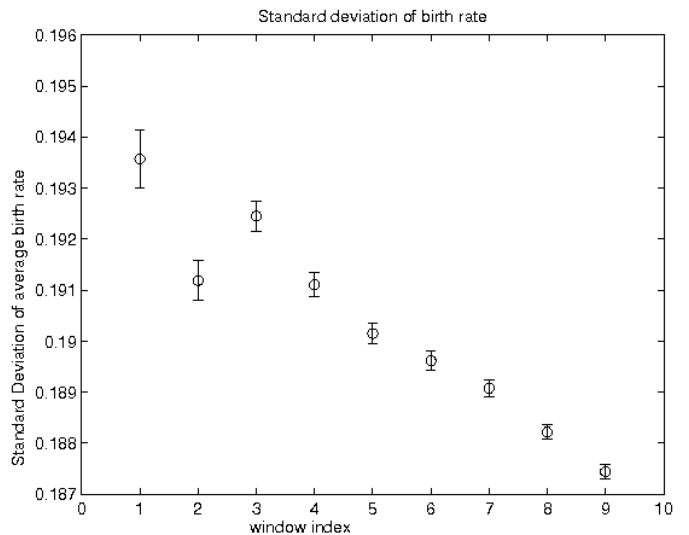}
\caption{The standard deviation of birth rate fluctuations decreases over time. The window index, $x$, gives the start and end points of each sample window using the following conversion: $t_{\small{start}}=10\cdot 2^x$ and $t_{\small{end}} = 18\cdot 2^x$. Error bars show that while the effect is small, it is significant.}
\label{fig:fluctuations}
\end{figure}

Figure~\ref{fig:fluctuations} shows the decrease in the standard deviation of these fluctuations plotted 
over logarithmically spaced time intervals. The fluctuations were only measured in the metastable parts of the dynamics.
 This suggests that the system evolves in such a way that barrier hopping becomes less and less likely, 
 not because of an increase in barrier height, but due to a decrease in the width of the fluctuations in the system.

This is not the end of the story, since these fluctuations are themselves governed by the nature of the network of interactions.
 Why the systems chooses configurations less susceptible to fluctuations  has not been answered in this paper, 
 and will form the focus of future work.

Acknowledgement: HJJ and DJ gratefully acknowledge the EPSRC for funding under grant EP/D051223

\end{document}